# Ranking with Confidence: A Probabilistic Framework for Deterministic Ranking Methods

**Shunpu Zhang**

**School of Data, Mathematical, and Statistical Sciences**

**University of Central Florida**

**Orlando, FL 32816**

Running head: Probabilistic Framework for Deterministic Rankings

ORCID Identifiers: 0000-0002-6999-7668

**Abstract**

Rankings are central to decision-making in fields ranging from education to online platforms, yet classical deterministic methods—such as the Borda count method or Copeland-type pairwise methods—ignore uncertainty due to sampling noise or incomplete data. We propose a probabilistic framework that treats true ranks as latent random variables, enabling quantification of ranking uncertainty. We introduce new ranking criteria based on pairwise dominance probabilities, derive approximate inference procedures, and provide a novel Worst-Best rank method to construct simultaneous and individual confidence intervals for ranks. Our approach is the first to provide formal uncertainty quantification for classical deterministic rankings. It is inherently robust to missing data: unlike Copeland-type methods, which penalize entities with fewer observed comparisons by assigning them fewer wins, our pairwise probability model adjusts for incompleteness, eliminating bias toward items with more complete records. The resulting rankings reflect underlying performance rather than data availability, enhancing fairness, transparency, and statistical reliability in high-stakes applications.

## 1. Introduction

Rankings are essential in sectors like sports, elections, education, and business for comparing performance and guiding decisions (Zhang et al., 2014). Methodologies for determining rankings generally fall into non-parametric count methods and parametric statistical models. Non-parametric methods, such as the Borda count (Borda, 1781) and Copeland pairwise comparison (Copeland, 1951), aggregate rankings without distributional assumptions. Conversely, parametric methods employ probabilistic frameworks, evolving from Thurstone's law of comparative judgment (Thurstone, 1927) and the Bradley–Terry model (Bradley and Terry, 1952) to the Plackett–Luce model (Plackett, 1975). Recent research has focused on asymptotic theories (Xie et al. (2009), Jang et al., 2018; Han and Xu, 2023; Fan et al., 2023, 2024a,b,c) and covariate-assisted ranking (Dong et al., 2025), among others.

While parametric methods are valuable in marketing and psychology, their complexity and reliance on assumptions limit their transparency in official elections. Non-parametric methods are deterministic and transparent but lack theoretical depth and measures of uncertainty, such as confidence intervals. To bridge this gap, this article introduces a non-parametric approach that incorporates latent random variables to model uncertainty. We develop statistical techniques to quantify ranking accuracy and stability by constructing confidence intervals for ranks obtained from deterministic ranking methods.

## 2. Probabilistic Ranking Framework for Deterministic Ranking

Consider the problem of ranking $N$ entities $E_1, \ldots, E_N$. A ranker is tasked with ordering these items under the assumption that larger values indicate better quality or performance. The goal is to construct a ranking list

$$E_{i_1} \prec E_{i_2} \prec \cdots \prec E_{i_N},$$

representing an ordered preference of $E_1, \ldots, E_N$ and "$\prec$" denotes "is less preferred than."

### Main Idea 1: Latent Variable Perspective

We associate the ranker's perception of $E_1, \ldots, E_N$ with latent random variables $X_1, \ldots, X_N$. Traditional parametric methods often rely on numerical summaries such as means to rank random variables. However, comparing means can be misleading because it ignores variability and the frequency of outperforming competitors. For example, consider ranking two restaurants $A$ and $B$. Let $X_A$ and $X_B$ be the latent random variables reflecting how customers rank these two restaurants and assume $X_A$ and $X_B$ follow the following distributions:

| Table 1 here |
|---|

Here, $E[X_A] = 1.70 < E[X_B] = 1.80$, yet $P(X_A \leq X_B) = .44$. This illustrates a counterintuitive scenario: a). Restaurant B has a higher expected (mean) rank due to rare exceptional outcomes. b). Restaurant A, while lower on average, outperforms Restaurant 2 in 56% of visits. This example demonstrates the value of probability-based ranking: ranking options by $P(X_A \leq X_B)$, the probability that restaurant *A* performs no better than restaurant *B*—captures the likelihood of outperforming alternatives in any single trial. This approach better aligns with real-world decision-making, where outcomes unfold one instance at a time, and consistent reliability often outweighs the allure of rare, exceptional results, which is especially valuable in high-stakes domains like elections, sports, stock trading, and strategic business decisions, where robustness, predictability, and fairness are paramount.

In a general setting with *N* latent random variables $X_1, \ldots, X_N$ associated with *N* entities. The preceding discussion shows that we can define the rank of $X_i$ based on the Pairwise Dominance Probabilities defined as follows:

**The Cumulative Pairwise Dominance Probabilities (CPDP) Criterion**

Define the sum of the Pairwise Dominance Probabilities for $X_i$ as

$$s_i = \sum_{k=1}^{N} P(X_k \leq X_i). \tag{1}$$

With available $s_i$'s, the rank of $X_i$ is obtained by ranking $s_i$ among $s_1, \ldots, s_N$:

$$s_i' = \sum_{k=1}^{N} I(s_k \leq s_i). \tag{2}$$

The rank defined in (2) uses rank 1 as the worst and $N$ as the best.

Define

$$R_i = \sum_{k=1}^{N} I(X_k \leq X_i), i = 1, \ldots, N, \tag{3}$$

where $I(\cdot)$ is the indicator function. We shall call $R_i$ the estimated rank of $X_i$ in the remainder of the paper. Note that

$$E[R_i] = \sum_{k=1}^{N} P(X_k \leq X_i) = s_i.$$

Hence, $R_i$ is an unbiased estimator of the sum of the Pairwise Dominance Probabilities $s_i$.

**The Cumulative Truncated Pairwise Dominance Probabilities (CTPDP) Criterion**

To account for the decisive nature of victory and defeat in tournaments and elections, we modify the above criterion by truncating probabilities at 1/2. If $P(X_k \leq X_i) > \frac{1}{2}$, we say $X_i$ wins against $X_k$. Define

$$t_i = \sum_{k=1}^{N} I[P(X_k \leq X_i) > 1/2]. \quad (4)$$

Unlike $s_i$'s in (2), which require further ranking, $t_i$'s in (4) provide ranks directly. As in (3), the rank defined in (4) uses rank 1 as the worst and $N$ as the best. To reverse this order (rank 1 as best), one can simply apply the following transformation:

$$t_i' = N + 1 - t_i, i = 1, \dots, N. \quad (5)$$

**Main Idea 2: Rank-Based Equivalence**

When only the estimated ranks $R_1, \dots, R_N$ defined in (1) are available, note the equivalence:

$$\{X_k \leq X_i\} \iff \{R_k \leq R_i\}, \forall k, i. \quad (6)$$

Thus, $s_i$ and $t_i$ defined in (1) and (4) have the following equivalent forms

$$s_i = \sum_{k=1}^{N} P(R_k \leq R_i), \quad (7)$$

and

$$t_i = \sum_{k=1}^{N} I[P(R_k \leq R_i) > 1/2]. \quad (8)$$

This equivalence allows all pairwise comparisons to be expressed solely in terms of ranks. It enables the construction of confidence intervals for ranks obtained under the CPDP and CTPDP criteria, which will be used extensively in later sections.

Ranking via pairwise probabilities compares items two at a time and aggregates their win/loss probabilities into an overall ordering. This approach is widely applied in machine learning, search engines, recommender systems, and sports analytics, as it remains reliable even with noisy or incomplete data. More generally, pairwise-probability–based ranking is closely related to stochastic dominance, probabilistic preference models (David, 1981; Felsenthal & Machover, 1997; Fishburn, 1974; Joe, 1990; Laslier, 1997), among others, and parametric frameworks such as the Bradley–Terry model. Unlike methods based on parametric models, which produces fully probabilistic rankings, our proposed methods generate deterministic ranks while embedding a probabilistic structure that supports confidence-interval–based uncertainty quantification. This provides clear ranking decisions together with rigorous measures of uncertainty, making the approach particularly suitable for applications that require both determinism and statistical reliability.

## 3. Estimation of Ranks by Aggregation

In real-world applications, there are often available multiple rankers (say *m*), i.e., for entity $i$, there are *m* observations $X_{i1}, X_{i2}, \dots, X_{im},\ i = 1, \dots, N$. We have the following latent data matrix

$$\boldsymbol{X} = \begin{bmatrix} X_{11} & \cdots & X_{1m} \\ \vdots & \vdots & \vdots \\ X_{N1} & \cdots & X_{Nm} \end{bmatrix}, \tag{9}$$

where each column is one replication of the latent random variables $X_1, \dots, X_N$, respectively. Since $\boldsymbol{X}$ is latent, we observe only the ranks of $X_1, \dots, X_N$. That is, instead of $X_{1l}, X_{2l}, \dots, X_{Nl}$, only their ranks $R_{1l}, R_{2l}, \dots, R_{Nl}, l = 1, \dots, m$, are available. Then we have available the following rank matrix

$$\boldsymbol{R} = \begin{bmatrix} R_{11} & \cdots & R_{1m} \\ \vdots & \vdots & \vdots \\ R_{N1} & \cdots & R_{Nm} \end{bmatrix} \tag{10}$$

where

$$R_{il} = \sum_{k=1}^{N} I(X_{kl} \le X_{il}), i = 1, \dots, N, \tag{11}$$

and $R_{il}$ is the estimated rank of $X_{il}$ in the *l*-th column of $\boldsymbol{X}$ obtained by using the rank function (3). The data matrix (9) is in general not available, but the introduction of it makes it possible to find the confidence intervals through the Rank-based equivalence property (6). The two quantities $s_i$ and $t_i$ defined by (1) and (4) require estimation of $P(X_k \le X_i), i, k = 1, \dots, N$. For convenience, we will denote $P(X_k \leq X_i)$ by $p_{ki}$ and its estimator by $\hat{p}_{ki}$. With (9) and (10), a natural estimator of $p_{ki}$ based on the rank function can be obtained by

$$\hat{p}_{ki} = \sum_{l=1}^{m} I(X_{kl} \leq X_{il})/m. \tag{12}$$

Then, an estimator of $s_i$ is

$$\begin{aligned}\hat{s}_i &= \sum_{k=1}^{N} \hat{p}_{ki} \\ &= \sum_{k=1}^{N} \sum_{l=1}^{m} I(X_{kl} \leq X_{il})/m \\ &= \sum_{l=1}^{m} \sum_{k=1}^{N} I(X_{kl} \leq X_{il})/m \\ &= \sum_{l=1}^{m} R_{il}/m \\ &= \bar{R}_i, \end{aligned} \tag{13}$$

where $R_{il} = \sum_{k=1}^{N} I(X_{kl} \leq X_{il})$ is the estimated rank of $X_{il}$ among the elements of the *l*-th column of the matrix (9) and $\bar{R}_i$ is the average of the $i_{th}$row of the rank matrix (10).

The estimated rank of $X_i$ can then obtained by

$$\begin{aligned}\hat{s}_i' &= \sum_{k=1}^{N} I(\hat{s}_k \leq \hat{s}_i), \\ &= \sum_{k=1}^{N} I(\bar{R}_k \leq \bar{R}_i). \end{aligned}$$

In summary, the estimated rank under the CPDP criterion can be simply conducted in the following two steps:

1. Calculate the row average (or the sum because they have the same denominator) of the rank matrix (10) and denote them by $\bar{r}_1, \ldots, \bar{r}_N$.
2. The rank of $\bar{r}_i$ in $\bar{r}_1, \ldots, \bar{r}_N$ is the rank of random variable $X_i$.

**Proposition 1:** For complete data, the method based on the CPDP Criterion is the Borda count method.

**Remark 1**: This is the first time in the literature which specifically point out that the Borda count method ranks according to the Cumulative Pairwise Dominance Probabilities (CPDP) Criterion.

**Remark 2**: The preceding discussion reveals that the Borda count method can equivalently be formulated within a pairwise comparison framework. The pairwise nature makes it possible to handle missing data because pairwise methods look at how each pair of candidates is ranked relative to each other by voters.

For the CTPDP Criterion, we have

$$\hat{t}_i = \sum_{k=1}^{N} I[\hat{p}_{ki} > 1/2]$$

$$= \sum_{k=1}^{N} I[\sum_{l=1}^{m} I(X_{kl} \leq X_{il})/m > 1/2]$$

$$= \sum_{k=1}^{N} I[c_{ki} > m/2] \tag{14}$$

where

$$c_{ki} = \sum_{l=1}^{m} I(X_{kl} \leq X_{il})$$

$$= \sum_{l=1}^{m} I(R_{kl} \leq R_{il}), \tag{15}$$

by the Rank-based equivalence (6) between $X_{kl} \leq X_{il}$ and $R_{kl} \leq R_{il}$.

**Proposition 2:** For complete data, the method based on the CTPDP Criterion is the pairwise method (the Copeland pairwise method). The only difference is that our method gives full rank for each entity in a tie while the Copeland pairwise method uses half rank for ties.

The proofs of Propositions 1 and 2 follow directly from the preceding discussion and are therefore omitted. For brevity, we shall call the methods based on the CPDP and CTPDP Criteria the CPDP method and the CTPDP method, respectively.

## 4. Confidence Intervals for the Ranks under the CPDP and CTPDP Criteria

### 4.1. Ranks under the CPDP criterion

Recall that the estimated ranks under the CPDP criterion is obtained through

$$\hat{s}_i' = \sum_{k=1}^{N} I(\hat{s}_k \leq \hat{s}_i) = \sum_{k=1}^{N} I(\bar{R}_k \leq \bar{R}_i).$$

It is not readily clear how to obtain the confidence interval of $s_i'$ from the above expression directly. However, the Ranked-based equivalence (6) makes it possible to get the confidence intervals from working on the latent data matrix (9), in which we assume the independence among the rows, the columns, and between the rows and columns (which is reasonable since the rankers and entities under comparison work independently- we shall call this the ranker and entity (R&E) independence assumption).

**Theorem 1:** Under the R&E independence assumption, we have the following results: For $i \in [1, N]$,

1. $\hat{s}_i$ is unbiased for $s_i$.
2. $V(\hat{s}_i) = \frac{1}{m}(\sum_{k=1}^{N} p_{ki}(1 - p_{ki}) + \sum_{1 \le s \ne t \le N} (p_{st(i)} - p_{si}\, p_{ti}),$ (16)

where $p_{ki} = P(X_k \le X_i),\ p_{st(i)} = P((X_s \le X_i) \cap (X_t \le X_i))$.

3. $\hat{s}_i \sim N\big(E(\hat{s}_i), V(\hat{s}_i)\big)$ asymptotically.

The proof of Theorem 1 is deferred to the appendix. Note that $p_{ki}$'s are unavailable in practice, but they can be estimated by

$$\hat{p}_{ki} = \sum_{l=1}^{m} I(X_{kl} \le X_{il})/m$$
$$= \sum_{l=1}^{m} I(R_{kl} \le R_{il})/m, \qquad (17)$$

and $p_{st(i)}$ can be estimated by

$$\hat{p}_{st(i)} = \sum_{l=1}^{m} I(R_{sl} \le R_{il}) I(R_{tl} \le R_{il})/m. \qquad (18)$$

Plug (17) and (18) in (16), we have an estimator of $V(\hat{s}_i)$ as follows

$$\hat{V}(\hat{s}_i) = \frac{1}{m}(\sum_{k=1}^{N} \hat{p}_{ki}(1 - \hat{p}_{ki}) + \sum_{1 \le s \ne t \le N} (\hat{p}_{st(i)} - \hat{p}_{si}\hat{p}_{ti})). \qquad (19)$$

With the asymptotic results obtained above, a level 100(1-$\alpha$)% asymptotic confidence interval for $s_i$ is provided by

$$\hat{s}_i \pm z_{\frac{\alpha}{2}} \sqrt{\hat{V}(\hat{s}_i)},\ i = 1, \ldots, N. \qquad (20)$$

The intervals obtained by (20) are the individual confidence intervals for the cumulative pairwise probabilities $s_i$'s, not for the ranks. It is pointed out in Zhang et al. (2014) that ranking is by nature simultaneous. Below we propose methods to construct the simultaneous confidence intervals of the ranks obtained based on the CPDP Criteria.

### 4.2. The Worst-Best Rank method

Assume that the level of the simultaneous confidence intervals for the ranks is set at 100(1-$\alpha$)%. Using the Bonferroni method, the level of significance for individual confidence intervals should be set at 100(1-$\alpha/N$)%. For 1≤ $i$≤ $N$, denote by $(L_i, U_i)$ the lower and upper bounds of the confidence interval for $s_i$ from (20) at level 100(1-$\alpha/N$)%. The lower bound for $s_i'$, the rank of $s_i$, can be obtained by comparing $L_i$ with the upper bounds of all other confidence intervals, i.e., with $U_j, j = 1, \ldots, N$. Such obtained rank of $s_i$ is the worst-case scenario (the lowest rank) of the rank of $s_i$. Similarly, the upper bound for $s_i'$, the rank of $s_i$, can be obtained by comparing the upper bound $U_i$, with the lower bounds of all other confidence intervals, i.e., with $L_j, j = 1, \ldots, N$. Such obtained upper bound for the rank of $s_i$ is the best-case scenario (the highest rank) of the rank of $s_i$. We shall call this method the Worst-Best Rank method.

**Proposition 3:** The level of the simultaneous confidence intervals obtained from the Worst-Best Rank method is at least 1-$\alpha$.

The proof of Proposition 3 is provided in the Appendix.

### 4.3. Constructing the simultaneous confidence intervals using the Worst-Best Rank method

Denote by $L_i$ and $U_i$ the lower and upper bounds of the level $100\left(1-\frac{\alpha}{N}\right)$%th percentiles of the asymptotic normal distribution $N\left(\hat{s}_i, \hat{V}(\hat{s}_i)\right)$. The simultaneous confidence intervals for each *i* can be obtained as follows:

a. Lower bound for $s_i'$: The rank of the lower bound $L_i$ of the level $100\left(1-\frac{\alpha}{N}\right)$% confidence interval for $s_i$ when compared with all upper bounds excluding *i*.

$WorstRank_i = 1 + \{j \neq i, U_j \leq L_i\}.$

b. Upper bound for $s_i'$: The rank of the upper bound $U_i$ of the level $100\left(1-\frac{\alpha}{N}\right)\%$ confidence interval for $s_i$ when compared with all lower bounds excluding *i*.

$BestRank_i = 1 + \{j \neq i, L_j \leq U_i\}.$

Then

$$Rank\ CI_i = [WorstRank_i, BestRank_i],\ i = 1, \dots, N \tag{21}$$

are the simultaneous level 100(1-$\alpha$)% confidence intervals for the ranks of the entities under the CPDP Criterion based on the Worst-Best (CPDP-WB) Rank method.

**4.4. Constructing the individual confidence intervals with the Worst-Best Rank method**

It is often of practical intertest of knowing the level 100(1-$\alpha$)% individual confidence interval for the ranks of the entities in comparison. Denote the individual confidence interval for $s_i$ of the *i*-th entity by $CI_i, i = 1, \dots, N$. With Theorem 1, the sampling distribution $N\left(\hat{s}_i, \hat{V}(\hat{s}_i)\right)$ was used in obtaining $CI_i$. Note that

$$\begin{aligned} P(CI_i) &\geq P\left(CI_i \cap_{1\leq l\neq i\leq N} CI_l\right) \\ &\geq 1 - P(\overline{CI}_i) - P(\cup_{1\leq l\neq i\leq N} \overline{CI}_l) \\ &\geq 1-\alpha, \end{aligned}$$

which can be achieved if $P(\overline{CI}_i) + P(\cup_{1\leq l\neq i\leq N} CI_l) \leq \alpha$, which is implied by

$$P(\overline{CI}_i) \leq \frac{k\alpha}{N} \text{ and } \sum_{1\leq l\neq i\leq N} P(\overline{CI}_l) \leq \frac{(N-k)\alpha}{N}, k = 1, \dots, N, \tag{22}$$

which is further implied by

$$P(\overline{CI}_i) \leq \frac{k\alpha}{N} \text{ and } P(\overline{CI}_l) \leq \frac{(N-k)\alpha}{N(N-1)}, 1 \leq l \neq i \leq N, k = 1, \dots, N. \tag{23}$$

If $k = 1$, (23) reduces to

$$P(\overline{CI}_i) \leq \frac{\alpha}{N} \text{ and } P(\overline{CI}_l) \leq \frac{\alpha}{N}, \quad 1 \leq l \neq i \leq N, \tag{24}$$

which are the level 100(1- $\alpha$)% Bonferroni simultaneous confidence intervals.

If $k = N - 1$, (23) is reduced to

$$P(\overline{CI}_i) \le \frac{(N-1)\alpha}{N} \text{ and for all} 1 \le l \ne i \le N,\ P(\overline{CI}_l) \le \frac{\alpha}{N(N-1)}. \quad (25)$$

For each $i$, we will find the value of $k$ in [1, $N$-1] which minimizes the length of $CI_i$, the level 100(1- $\frac{k\alpha}{N}$)% confidence interval for $s_i$. For this $k$, the confidence intervals ($CI_j$) for other entities at level 100(1-$\frac{(N-k)\alpha}{N(N-1)}$)% are obtained from $N\left(\hat{s}_j, \hat{V}\left(\hat{s}_j\right)\right), j \ne i$. Then, the level 100(1-$\alpha$)% individual confidence interval for the CPDP rank of the $i$-th entity $E_i$ will be constructed by applying the Worst-Best Rank method (i.e., the CPDP-WB Rank method) on these $CI_i$ and $CI_j, j \ne i$.

### 4.5. Confidence Intervals for the ranks under the CTPDP Criterion

We first derive the counterpart of Theorem 1 for the estimated rank derived under the CTPDP criterion.

**Theorem 2:** Under the R&E independence assumption, for $\hat{t}_i$ defined in (14), for $i$=1,…, $N$, we have

1. $E(\hat{t}_i) = \sum_{k=1}^{N} \sum_{s=\lceil mq \rceil}^{m} \binom{m}{s} (p_{ki})^s (1-p_{ki})^{m-s}$ (26)

2. The variance is bounded by

$$V(\hat{t}_i) \le \sum_{k=1}^{N} P(N_{k,i} > mq)(1 - P(N_{k,i} > m/2))$$
$$+2 \sum_{1 \le s < t \le N} [\, P\left(N_{s,i} > mq\right) \vee P\left(N_{t,i} > mq\right)$$
$$-P(N_{s,i} > mq)P(N_{t,i} > mq)], \quad (27)$$

where

$$\mathrm{P}\left(N_{j,i} > mq\right) = \sum_{s=\lceil mq \rceil}^{m} \binom{m}{s} \left(p_{ji}\right)^s \left(1-p_{ji}\right)^{m-s}, j = 1, \dots, N. \quad (28)$$

**Remark 3**. Equation (26) means that $\hat{t}_i$ is an unbiased estimator of $\sum_{k=1}^{N} P[\hat{P}(X_k \leq X_i) > 1/2]$ instead of $t_i = \sum_{k=1}^{N} I[P(X_k \leq X_i) > 1/2]$ defined in (8).

By plugging $\hat{p}_{ji}$ defined in (17) into (28), we obtain an estimate of $\mathrm{P}\big(N_{j,i} > mq\big)$, substituting this result into (27) gives the estimated upper bound for the variance $V(\hat{t}_i)$), denoted by $\hat{V}(\hat{t}_i)$. Recall that

$\hat{t}_i = \sum_{k=1}^{N} I[\hat{p}_{ki} > 1/2]$, and $\hat{p}_{ki} = \sum_{l=1}^{m} I(X_{kl} \leq X_{il})/m$.

Clearly, $\hat{t}_i$ is the sum of $N$ dependent Bernoulli random variables with success probabilities $P(\hat{p}_{ki} > 1/2)$, $k = 1, \ldots, N$. Thus, the Central Limit Theorem generally fails for $\hat{t}_i$ . Despite this non-normality and the aforementioned bias of $\hat{t}_i$ as an estimator for $t_i$, we derive confidence intervals for the CTPDP rank $t_i$ based on $N(\hat{t}_i, \hat{V}(\hat{t}_i))$ and employ the Worst-Best Rank method (Subsection 4.2) for simultaneous and individual inference. Note that (27) provides an upper bound for $V(\hat{t}_i)$, we anticipate that the enlarged variance will alleviate the issues caused by the biased estimate of $t_i$ and the non-normality of $\hat{t}_i$, and provides meaningful confidence intervals for the ranks $t_i$. Section 6 evaluates this performance via simulation.

## 5. Ranking with Missing Data

In ranked-choice voting, "bottom-imputation"—treating unranked candidates as tied for last—is a common practice to handle missing votes. However, this approach often distorts voter intent by interpreting neutrality or lack of knowledge as disapproval. In Borda Count systems, this penalizes unranked candidates and encourages strategic truncation. In Condorcet methods, such as the Schulze method (Schulze, 2011) and the Copeland pairwise method, it can artificially inflate losses in head-to-head matchups, potentially preventing the election of a true Condorcet winner.

Consequently, bottom-imputation is an inappropriate method for handling missing data. While the pairwise nature of Criteria (2) and (4) allows for the exclusion of unranked pairs, classical pairwise methods like Copeland remain susceptible to bias. In these systems, candidates with missing data suffer from fewer comparison opportunities, lowering their aggregate win counts regardless of actual performance. As will be demonstrated, the probabilistic nature of our proposed criteria effectively circumvents this bias.

To accommodate cases with missing data, we denote by $M_l$ the set of the row indices of the missing observations in the $l$-th column of the data matrix $\boldsymbol{X}$. Let $R_{il}$ be the rank of a non-missing $X_{il}$ among the non-missing elements of the $l$-th column of $\boldsymbol{X}$. Then, the rank of the non-missing $X_{il}$ among the non-missing elements of the $l$-th column is calculated as follows:

$$R_{il} = \sum_{k=1,\notin M_l}^{N} I(X_{kl} \le X_{il}), i = 1, \dots, N; l = 1, \dots, m. \quad (29)$$

For those with missing observations, their ranks will be missing and be denoted by N/A. For simplicity of notations, the resulting rank matrix is again denoted by $\boldsymbol{R}$ as in (10).

Denote the set of the columns in $\boldsymbol{X}$ with at least one missing observation in the $k$-th or $i$-th row by $M_{ki}$, a natural estimator of $p_{ki} = P(X_k \le X_i)$,

$$\hat{p}_{ki} = \sum_{l=1,l\notin M_{ki}}^{m} I(X_{kl} \le X_{il})/m_{ki}, k, i = 1, \dots, N. \quad (30)$$

where $m_{ki} = \sum_{l=1}^{m} I[l \notin M_{ki}]$, the number of columns without missing data on both $k$-th and $i$-th rows. Hence, an estimator of $s_i$ defined in (2) is

$$\begin{aligned}\hat{s}_i &= \sum_{k=1}^{N} \hat{p}_{ki} \\ &= \sum_{k=1}^{N} \sum_{l=1,l\notin M_{ki}}^{m} I(X_{kl} \le X_{il})/m_{ki}. \\ &= \sum_{k=1}^{N} \sum_{l=1,l\notin M_{ki}}^{m} I(R_{kl} \le R_{il})/m_{ki}. \quad (31)\end{aligned}$$

The last equality is due to the Rank based equivalence (6). The estimated rank of $X_i$ under the CPDP criterion is then obtained by

$$\hat{s}_i' = \sum_{k=1}^{N} I(\hat{s}_k \le \hat{s}_i), \quad (32)$$

Similarly, an estimator of $t_i$ defined in (4) is

$$\begin{aligned}\hat{t}_i &= \sum_{k=1}^{N} I[\hat{P}(X_k \le X_i) > 1/2] \\ &= \sum_{k=1}^{N} I[\sum_{l=1,l\notin M_{ki}}^{m} I(X_{kl} \le X_{il}) > m_{ki}/2] \\ &= \sum_{k=1}^{N} I[\sum_{l=1,l\notin M_{ki}}^{m} I(R_{kl} \le R_{il}) > m_{ki}/2], \quad (33)\end{aligned}$$

and (33) is the estimated rank of the $i$-th entity under the CTPDP Criterion.

**Remark 4**. When there are missing data, the ranks obtained from the CPDP method (i.e., calculated from (32)) will be different from the Borda count method and those obtained from the CTPDP method (i.e., calculated from (33)) will be different from the Copeland pairwise method because the CPDP and CTPDP methods are based on the pairwise dominance probabilities while the Borda and Copeland pairwise methods are based on the aggregated counts of winnings.

### 5.1. Confidence Intervals for the Ranks under the CPDP Criterion

**Theorem 3**. Under the R&E independence assumption, we have the following results:

For $i = 1, \ldots, N$,

1. $\hat{s}_i$ is unbiased for $s_i$

2. $V(\hat{s}_i) = \sum_{k=1}^{N} \frac{1}{m_{ki}} p_{ki}(1 - p_{ki}) + \sum_{1 \le s \ne t \le N} \frac{1}{m_{si} m_{ti}} (|M_{si}^C \cap M_{ti}^C| ( p_{st(i)} - p_{si}\, p_{ti}),$ (34)

where $p_{ki} = P(X_k \le X_i)$ and $p_{st(i)} = P((X_s \le X_i) \cap (X_t \le X_i)$, $M_{si}^C$, $M_{ti}^C$ are the complementary sets of $M_{si}$ and $M_{ti}$, respectively and the notation $|\cdot|$ indicates the cardinality of a matrix. The proof of Theorem 3 is deferred to Supplemental Materials.

3. $\hat{s}_i \sim N\big(E(\hat{s}_i), V(\hat{s}_i)\big)$ asymptotically.

Note that $p_{ki}$'s in (34) can be estimated by (30) and $p_{st(i)}$ can be estimated by

$$\hat{p}_{st(i)} = \sum_{l \in M_{si}^C \cap M_{ti}^C} I(X_{sl} \le X_{il}) I(X_{tl} \le X_{il}) / m_{st(i)},$$

where $m_{st(i)} = \sum_{l=1}^{m} I\big[l \in M_{si}^C \cap M_{ti}^C\big]$. Hence, the estimated variance $V(\hat{s}_i)$ is

$$\hat{V}(\hat{s}_i) = \sum_{k=1}^{N} \frac{1}{m_{ki}} \hat{p}_{ki}(1 - \hat{p}_{ki}) + \sum_{1 \le s \ne t \le N} \frac{1}{m_{si} m_{ti}} |M_{si}^C \cap M_{ti}^C| \big( \hat{p}_{st(i)} - \hat{p}_{si}\, \hat{p}_{ti}\big). \tag{35}$$

With (35), simultaneous and individual confidence intervals for the ranks under the CPDP criterion can be constructed using the Worst-Best Rank method discussed in Section 4.

### 5.2. Confidence Intervals for the Ranks under the CTPDP Criterion

**Theorem 4**. Under the R&E independence assumption, we have the following results.

1. $E(\hat{t}_i) = \sum_{k=1}^{N} P(\, N^*_{k,i} > \frac{m_{ki}}{2}),$ (36)

where $N^*_{k,i} = \sum_{l \notin M_{ki}}^{m} I(\, X_{kl} \le X_{il}).$

2. $V(\hat{t}_i) \le \sum_{k=1}^{N} P\left(N^*_{k,i} > m_{ki}q\right)\left(1 - P\left(N^*_{k,i} > m_{ki}q\right)\right)$

$+ \sum_{1 \le s \ne t \le N}(P\left(N^*_{s,i} > m_{ki}q\right) \vee P\left(N^*_{t,i} > m_{ki}q\right) - P\left(N^*_{s,i} > mq\right)P\left(N^*_{t,i} > mq\right)),$ (37)

where $P\left(N^*_{k,i} > m_{ki}q\right) = \sum_{s=\lceil m_{ki}q \rceil}^{m_{ki}} \binom{m_{ki}}{s} (p_{ki})^s \,(1 - p_{ki})^{m_{ki}-s}.$ (38)

Theorems 3 and 4 can be proved along the same lines as those of Theorems 1 and 2 and hence is omitted. Plug in the estimate of $p_{ki}$ by $\hat{p}_{ki}$ defined in (30) in $P\left(N^*_{k,i} > m_{ki}q\right)$ of (38), and subsequently substitute it to (37), we can obtain an estimate of the upper bound for $V(\hat{t}_i)$, denote it by $\hat{V}(\hat{t}_i)$.

As indicated by Theorem 2 and part (1) of Theorem 4, $\hat{t}_i$ is an unbiased estimator of $\sum_{k=1}^{N} P[\hat{P}(\, X_k \le X_i) > 1/2]$ instead of the quantity of interest $t_i = \sum_{k=1}^{N} I[P(\, X_k \le X_i) > 1/2]$. Furthermore, $\hat{t}_i$ does not satisfy the conditions required for the Central Limit Theorem. Despite these limitations, we construct simultaneous and individual confidence intervals for $t_i$ using the normal approximation $N\left(\hat{t}_i, \hat{V}(\hat{t}_i)\right)$ combined with the Worst-Best Rank method we proposed in Section 4. The performance of these intervals is evaluated numerically in Section 6.

## 6. Simulations

In this section we use simulations to evaluate the performance (the coverage probability) of the CPDP and CTPDP methods. We generated the data matrix $\boldsymbol{X}$ defined by (9) from the following independent random variables $X_1, X_2, \ldots, X_N$ with

$$\boxed{X_{ij} \sim \mathcal{N}(\mu_i, \sigma_i^2)}, i = 1, \ldots, N; j = 1, \ldots, m.$$

Then, the rank matrix $\boldsymbol{R}$ defined by (10) is formed using (11).

**Case 1:** $\boxed{X_i \sim \mathcal{N}(\mu_i, \sigma_i^2)}, \mu_i = i, \sigma_i^2 = 1, i = 1, \ldots, N(= 10)$. In this case, the theoretical CPDP and CTPDP ranks are the same as those of the means.

**Case 2:** Same as Case 1 with 40% of rows missing in the rank matrix ***R*** defined in (10) and up to 40% missing observation in each row of ***R*** with missing data.

**Case 3:** $\boxed{X_i \sim \mathcal{N}(\mu_i, \sigma_i^2)}, \mu_i = i, \sigma_i^2, i = 1, \ldots, N(= 10)$. In this case, the CPDP ranks are different from those of the theoretical means, CTPDP ranks, and are reported in the following table.

| Table 2 here |
|---|

**Case 4**: Same as Case 3 with 40% of rows missing in the rank matrix ***R*** defined by (10) and up to 30% missing observation in each row of ***R*** with missing data.

Empirical simultaneous coverage probabilities were evaluated for both CPDP and CTPDP methods with the Worst-Best rank Method across sample sizes ranging from m = 5 to 55. As demonstrated in Figure 1, the CPDP-WB Rank method exhibits good control of simultaneous coverage probabilities when $m \geq 10$ (=$N$). For $m < N$, satisfactory coverage performance is observed only in Case 1. However, in scenarios involving missing data (Cases 2 and 4) or discrepancies between theoretical CTPDP ranks and CPDP/mean ranks (Case 3), CPDP-WB method's coverage probabilities consistently fall below the nominal level when $m$ < N. In all four cases, the CTPDP with the Worst-Best Rank (CTPDP-WB) method has satisfactory control of the confidence level. We observe a systematic over-coverage phenomenon where empirical probabilities exceed the nominal 95% level. The over-coverage is more severe for the CTPDP W-B Rank method, partially caused the enlarged variance estimate (37). Nevertheless, this over-coverage phenomenon, replicated across all simulations, appears intrinsic to confidence intervals for ranks due to the discrete nature of the underlying data.

Figure 1. Empirical coverage probabilities of Simultaneous Confidence Intervals of CPDP W-B and CTPDP W-B methods

| Figure 1 here |
|---|

Figures 2 and 3 present the empirical coverage probabilities for individual confidence intervals under the CPDP W-B and CTPDP W-B methods, respectively. These results corroborate the findings observed in Figure 1.

Figure 2. Empirical coverage probabilities of Individual Confidence Intervals of the CPDP W-B method

Figure 2 here

Figure 3. Empirical coverage probabilities of Individual Confidence Intervals of the CTPDP W-B method

Figure 3 here

As discussed in Section 4, the CTPDP estimator $\hat{t}_i$ may not follow an asymptotic normal distribution. Nevertheless, the simulation results indicate that the rank confidence intervals derived from the CTPDP W-B method maintain satisfactory control over the confidence level, albeit conservatively. This conservativeness likely stems from the variance estimate $\hat{V}(\hat{t}_i)$ we used is an upper bound for $V(\hat{t}_i)$, rather than its exact value, which yields wider confidence intervals and higher coverage probabilities. Finally, while excessive conservativeness can theoretically obscure rank distinguishability, our extensive simulations confirm that both CPDP W-B Rank and CTPDP W-B Rank methods efficiently separate ranks in all four cases. These specific results are omitted due to space constraints.

## 7. Real data Case: NFL Quarterback Ranking data

Table 3 shows the ranking lists of the NFL starting quarterbacks from 13 experts in week 12 of season 2014. The table was provided and studied in Yi et al. (2019) in which they used covariates (some summary statistics of the NFL players) to assist their study.

Table 3 here

For this data set, the number of experts (*m*=13) is smaller than the number of players (*N*=24). In Table 4, we reported the ranks obtained from the proposed CPDP W-B and CTPDP W-B methods. As a comparison, we also included the aggregated ranks from the Bayesian Analysis of Rank data with entities' Covariates and rankers' (unknown) Weights (BARCW ) method (Yi et al. (2019) in which the authors concluded that BARCW seems to be the most appropriate model for the NFL ranking data among the methods discussed in their paper. We also included the ranks from the Borda count method reported in Yi et al. (2019). It is unclear how the missing data were handled from their paper.

Table 4 here

We calculated the sum of squared errors (SSE) between the estimated ranks from each method and the ranks from each expert. For those experts with missing ranks, we adjusted the estimated ranks with respect to the missing ranks. The SSE values are as follows.

Table 5 here

Table 5 shows that our proposed CPDP W-B Rank method has the smallest SSE among all four methods, followed by BARCW and the CTPDP W-B Rank method. The Borda count method from Yi et al. (2019) has the largest SSE. It is understandable that the CTPDP W-B Rank method has larger SSE than the CPDP W-B method because it has coarser resolution than the CPDP W-B Rank method caused by the truncation on the pairwise dominance probability.

Figure 4 displays the simultaneous and individual confidence intervals for player ranks derived from the CPDP W-B method (Panels A and B) and the CTPDP W-B method (Panels C and D), respectively. In Panel A, the simultaneous intervals (95% confidence level) reveal distinct performance tiers among the quarterbacks. Andrew Luck stands out as the elite performer with an interval of [1, 2]; while statistically indistinguishable from Aaron Rodgers ([1, 3]) and Peyton Manning ([2, 6]), Luck is significantly separated from the rest of the cohort. In contrast, most quarterbacks fall into an indistinguishable "middle tier" characterized by wide, overlapping intervals. For instance, although Tom Brady (Rank 4, Interval [3, 8]) is ranked higher than Tony Romo (Rank 5, Interval [4, 8]), their performance levels are statistically indistinguishable due to the overlap. However, these two players are collectively distinct from lower-ranked players (e.g., Russell Wilson, starting at Rank 9). Panels B and D of Figure 4 (individual intervals) offer sharper separation, showing Luck at [1, 1] and Rodgers/Manning at [2, 3]. However, we avoid claiming statistical significance based on these individual intervals. As ranking is inherently a simultaneous inference problem, rigorous claims regarding group differences require simultaneous intervals. Finally, Panels C and D (CTPDP W-B method) show less differentiation among players compared to Panels A and B (CPDP W-B Method). This reduced resolution is due to the truncation of pairwise dominance probabilities in the CTPDP criterion, which results in wider margins of error and effectively blurs the distinctions between rankings.

**Figure 4.** Confidence Intervals from CPDP W-B Rank Method (Panels A and B) and CTPDP W-B Method (Panels C and D) for ranks of NFL starting quarterbacks

Figure 4 here

## 8. Conclusion and Discussion

This paper introduces a probabilistic paradigm shift in deterministic ranking methodology. By treating true ranks as latent random variables, our framework quantifies uncertainty arising from sampling noise and incomplete data—dimensions often overlooked by traditional deterministic methods like Borda or Copeland. We introduce novel criteria based on pairwise dominance probabilities and the Worst-Best rank method, which constructs simultaneous and individual confidence intervals for ranks to enhance decision-making transparency.

Our framework is inherently robust to data incompleteness. Unlike classical deterministic methods that bias rankings against entities with fewer comparisons, our model employs probabilistic adjustments to ensure rankings reflect actual performance rather than data availability. We further delineate the appropriate use of our criteria: CTPDP is optimal for identifying aggregate winners where margins are irrelevant, whereas CPDP should be used when incorporating the magnitude of dominance is necessary. Future research will extend these methodologies to handle complex dependent data structures, such as longitudinal ranking datasets.

**APPENDIX**

**Proof of Proposition 3:**

$$P(\cap_{i=1}^{N} s_i' \in (L_i, U_i)) = 1 - P(\cup_{i=1}^{N} s_i' \notin (L_i, U_i))$$

$$\geq 1 - \sum_{i=1}^{N} P(s_i' \notin (L_i, U_i))$$

$$=1\text{-}N\left(\frac{\alpha}{N}\right)$$

$$=1\text{-}\alpha.$$

Note that $\cap_{i=1}^{N} s_i' \in (L_i, U_i)$ implies the rank $s_i'$ of $X_i$ falls between $1 + \{j \neq i, U_j \leq L_i\}$ and $1 + \{j \neq i, L_j \leq U_i\}$ for each $1 \leq i \leq N$. The proof is complete.

**Proof of Theorem 1:**

(1). Unbiasedness

$$E(\hat{s}_i) = E\left(\frac{\sum_{l=1}^{m} \sum_{k=1}^{N} I(X_{kl} \leq X_{il})}{m}\right)$$

$$= \frac{\sum_{l=1}^{m} \sum_{k=1}^{N} P(X_{kl} \leq X_{il})}{m}$$

$$= \sum_{k=1}^{N} \sum_{l=1}^{m} P(X_{kl} \leq X_{il})/m$$

$$= \sum_{k=1}^{N} P(X_k \leq X_i)$$

$$=s_i,$$

since $X_{j1}, \dots, X_{jm}$ are i.i.d, denote it by $X_j$ for $j = 1, \dots, N$. So, $\hat{s}_i$ is unbiased for $s_i$.

(2). Variance of the ranks under the CPDP criterion

$$V(\hat{s}_i) = V\left(\frac{\sum_{l=1}^{m} \sum_{k=1}^{N} I(X_{kl} \leq X_{il})}{m}\right)$$

$$= V\left(\frac{\sum_{k=1}^{N} \sum_{l=1}^{m} I(X_{kl} \leq X_{il})}{m}\right)$$

$$= \frac{1}{m^2} V\left(\sum_{k=1}^{N} N_{k,i}\right), \tag{A1}$$

where $N_{k,i} = \sum_{l=1}^{m} I(X_{kl} \le X_{il})$. Note that $N_{k,i}$ follows a binomial distribution $Binom(m,\ p_{ki})$, where $p_{ki} = P(X_k \le X_i)$. We have $E\big(N_{k,i}\big) = m\, p_{ki}$ and $V\big(N_{k,i}\big) = m\, p_{ki}(1-\ p_{ki})$.

Since

$$V\big(\textstyle\sum_{k=1}^{N} N_{k,i}\big) = \sum_{k=1}^{N} V\big(N_{k,i}\big) + \sum_{1\le s\ne t\le N} cov\big(N_{s,i}, N_{t,i}\big), \quad \text{(A2)}$$

and further note that

$$cov\big(N_{s,i}, N_{t,i}\big) = E(N_{s,i}N_{t,i}) - E(N_{s,i})E(N_{t,i})$$
$$= E\big(N_{s,i}N_{t,i}\big) - m^2\, p_{s<i}\, p_{t<i}$$

and

$$E\big(N_{s,i}N_{t,i}\big) = E\big((\textstyle\sum_{l=1}^{m} I(X_{sl} \le X_{il})(\sum_{l=1}^{m} I(X_{tl} \le X_{il})\big)$$
$$= E(\textstyle\sum_{l=1}^{m} I(X_{sl} \le X_{il}) I(X_{tl} \le X_{il})$$
$$+ \textstyle\sum_{1\le l_1\ne l_2\le m} I(X_{sl_1} \le X_{il_1})\, I(X_{tl_2} \le X_{il_2})$$
$$= \textstyle\sum_{l=1}^{m} P((X_{sl} \le X_{il})\ \cap (X_{tl} \le X_{il}))$$
$$+ \textstyle\sum_{1\le l_1\ne l_2\le m} P((X_{sl_1} \le X_{il_1}) \cap (X_{tl_2} \le X_{il_2}))$$
$$= \textstyle\sum_{l=1}^{m} P((X_s \le X_i)\ \cap (X_t \le X_i)) + \sum_{1\le l_1\ne l_2\le m} P(X_s \le X_i)\, P\,(X_t \le X_i)$$
$$= mP((X_s \le X_i) \cap (X_t \le X_i) + m(m-1)P(X_s \le X_i)P(X_t \le X_i)$$
$$= m\, p_{st(i)} + m(m-1)p_{si}\, p_{ti},$$

where $p_{st(i)} = P((X_s \le X_i) \cap (X_t \le X_i)$. Hence,

$$cov(N_{s,i}, N_{t,i}) = E(N_{s,i}N_{t,i}) - E(N_{s,i})E(N_{t,i})$$
$$= E\big(N_{s,i}N_{t,i}\big) - m^2\, p_{s<i}\, p_{t<i}$$
$$= m\, p_{st(i)} + m(m-1)p_{si}\, p_{ti} - m^2\, p_{si}\, p_{ti}$$
$$= m(p_{st(i)} - p_{si}\, p_{ti}). \quad \text{(A3)}$$

Combine (A2) and (A3), we have

$$V\left(\sum_{k=1}^{N} N_{k,i}\right) = \sum_{k=1}^{N} V\big(N_{k,i}\big) + \sum_{1\le s\ne t\le N} cov\big(N_{s,i}, N_{t,i}\big)$$
$$= m\textstyle\sum_{k=1}^{N}\ p_{ki}(1-\ p_{ki}) + \sum_{1\le s\ne t\le N}\ (p_{st(i)} - p_{si}\, p_{ti})).$$

Hence,

$$V(\hat{s}_i) = \frac{1}{m}(\sum_{k=1}^{N} p_{ki}(1 - p_{ki}) + \sum_{1 \le s \ne t \le N} (p_{st(i)} - p_{si}\, p_{ti})).$$

The proof is complete.

(3). The asymptotic normality of $\sum_{l=1}^{m} \sum_{k=1}^{N} I(X_{kl} \le X_{il})$ follows from the following facts:

(i). under the R&E independence assumption, $U_l \triangleq \sum_{k=1}^{N} I(X_{kl} \le X_{il})$ is the sum of *N* dependent Bernoulli random variables with probability of success $\mathrm{P}(X_k \le X_i)$.

(ii). $U_l$'s are i.i.d. with mean $m_i = \sum_{k=1}^{N} P(X_k \le X_i)$ and variance $Var(\sum_{k=1}^{N} I(X_{kl} \le X_{il})), l =$ 1, ..., *N*.

**Proof of Theorem 2:**

(i).

$$E(\hat{t}_i) = \sum_{k=1}^{N} P(\hat{P}(X_k \le X_i) > 1/2)$$

$$= \sum_{k=1}^{N} P\left(\sum_{l=1}^{m} I(X_{kl} \le X_{il}) > m/2\right)$$

$$= \sum_{k=1}^{N} P(N_{k,i} > m/2),$$

where $N_{k,i} = \sum_{l=1}^{m} I(X_{kl} \le X_{il})$.

Note that $N_{k,i}$ follows a binomial distribution $Binom(n, p)$, where $p = P(X_k \le X_i)$. We have

$$\mathrm{P}\left(N_{k,i} > \frac{m}{2}\right) = \sum_{s=\left\lceil \frac{m}{2} \right\rceil}^{m} \binom{m}{s} (p_{ki})^s \, (1 - p_{ki})^{m-s}, \tag{A4}$$

and

$$E(\hat{t}_i) = \sum_{k=1}^{N} \sum_{s=\left\lceil \frac{m}{2} \right\rceil}^{m} \binom{m}{s} (p_{ki})^s \, (1 - p_{ki})^{m-s}. \tag{A5}$$

(ii). Variance of the CTPDP method

$$V(\hat{t}_i) = V\left(\sum_{k=1}^{N} I\left(\sum_{l=1}^{m} I(X_{kl} \le X_{il}) > m/2\right)\right)$$

$$= V(\sum_{k=1}^{N} I(N_{k,i} > m/2))$$

$$= \sum_{k=1}^{N} V(I(N_{k,i} > m/2)) + 2 \sum_{1 \le s < t \le N} cov\big(I(N_{s,i} > mq), I(N_{t,i} > m/2)\big)$$

$$= \sum_{k=1}^{N} P(N_{k,i} > m/2)(1 - P(N_{k,i} > m/2))$$

$$+2 \textstyle\sum_{1 \le s < t \le N} cov\left(I(N_{s,i} > \frac{m}{2}\right), I(N_{t,i} > m/2)). \qquad \text{(A6)}$$

It can be seen that

$$cov\left(I(N_{s,i} > \frac{m}{2}\right), I(N_{t,i} > m/2))$$

$$= E\left(I(N_{s,i} > \frac{m}{2}\right) I(N_{t,i} > m/2)) - E(I(N_{s,i} > m/2) E(I(N_{s,i} > m/2))$$

$$= E\left(I(N_{s,i} > \frac{m}{2}\right) I(N_{t,i} > m/2)) - P(N_{s,i} > m/2) P(N_{s,i} > m/2). \qquad \text{(A7)}$$

Since $I(N_{t,i} > m/2) I(N_{t,i} > m/2) \le I(N_{s,i} > m/2) \vee I(N_{t,i} > m/2)$, we have

$$E\left(I(N_{s,i} > \frac{m}{2}\right) I(N_{t,i} > m/2)) \le E\left(I(N_{s,i} > \frac{m}{2}\right) \vee E\left(I(N_{t,i} > \frac{m}{2}\right)$$

$$= P\left(I(N_{s,i} > \frac{m}{2}\right) \vee P\left(I(N_{t,i} > \frac{m}{2}\right)$$

and

$$cov\left(I(N_{s,i} > \frac{m}{2}\right), I(N_{t,i} > m/2))$$

$$\le P\left(I(N_{s,i} > \frac{m}{2}\right) \vee P\left(I(N_{t,i} > \frac{m}{2}\right) - P(N_{s,i} > m/2) P(N_{s,i} > m/2). \qquad \text{(A8)}$$

Plug (A8) in (A7) and combine with (A6), we have

$$V(\hat{t}_i) = V\left(\sum_{k=1}^{N} I\left(\sum_{l=1}^{m} I(X_{kl} \le X_{il}\right) > \frac{m}{2}\right)$$

$$\le \sum_{k=1}^{N} P(N_{k,i} > m/2)(1 - P(N_{k,i} > m/2))$$

$$+2 \textstyle\sum_{1 \le s < t \le N} ( P\left(I(N_{s,i} > \frac{m}{2}\right) \vee P\left(I(N_{t,i} > \frac{m}{2}\right)$$

$$-P(N_{s,i} > m/2)P(N_{s,i} > m/2)).$$

The proof is complete.

Table 1. Probability distribution of $X_A$ and $X_B$

| $x$ | 1 | 2 | 3 | 4 | 5 |
|---|---|---|---|---|---|
| $P(X_A = x)$ | 0.30 | 0.70 | 0.00 | 0.00 | 0.00 |
| $P(X_B = x)$ | 0.80 | 0.00 | 0.00 | 0.00 | 0.20 |

Table 2. The means, CPDP, CTPDP values and their ranks

| | $\mu_i$ | $\sigma_i^2$ | CPDP | CPDP Ranks | CTPDP | CTDPD Ranks |
|---|---|---|---|---|---|---|
| $X_1$ | 1 | 9 | 2.0547 | 2 | 1 | 1 |
| $X_2$ | 2 | 1 | 1.9882 | 1 | 2 | 2 |
| $X_3$ | 3 | 1 | 2.8794 | 3 | 3 | 3 |
| $X_4$ | 4 | 1 | 3.9209 | 4 | 4 | 4 |
| $X_5$ | 5 | 1 | 5.0120 | 5 | 5 | 5 |
| $X_6$ | 6 | 1 | 6.1067 | 6 | 6 | 6 |
| $X_7$ | 7 | 1 | 7.1850 | 7 | 7 | 7 |
| $X_8$ | 8 | 1 | 8.2022 | 8 | 8 | 8 |
| $X_9$ | 9 | 1 | 9.0606 | 10 | 9 | 9 |
| $X_{10}$ | 10 | 16 | 8.5903 | 9 | 10 | 10 |

Table 3. Ranking lists of NFL starting quarterbacks from 13 different experts

| Player | $\tau_1$ | $\tau_2$ | $\tau_3$ | $\tau_4$ | $\tau_5$ | $\tau_6$ | $\tau_7$ | $\tau_8$ | $\tau_9$ | $\tau_{10}$ | $\tau_{11}$ | $\tau_{12}$ | $\tau_{13}$ |
|---|---|---|---|---|---|---|---|---|---|---|---|---|---|
| Andrew Luck | 1 | 1 | 1 | 3 | 3 | 1 | 1 | 1 | 1 | 1 | 1 | 1 | 1 |
| Aaron Rodgers | 2 | 3 | 4 | 2 | 1 | 2 | 3 | 3 | 2 | 2 | 3 | 4 | 3 |
| Peyton Manning | 3 | 2 | 5 | 4 | 2 | 3 | 2 | 2 | 3 | 4 | 4 | 2 | 2 |
| Tom Brady | 4 | 7 | 3 | 5 | 4 | 5 | 4 | 6 | 4 | 3 | 6 | 8 | 4 |
| Tony Romo | 9 | 5 | 6 | 1 | 5 | 4 | 5 | 4 | 5 | 5 | 7 | 6 | 6 |
| Drew Brees | 10 | 4 | 2 | 8 | 9 | 7 | 7 | 5 | 7 | 6 | 2 | 3 | 5 |
| Ben Roethlisberger | 6 | 8 | 7 | 7 | 7 | 6 | 6 | 10 | 6 | 7 | 5 | 7 | 7 |
| Ryan Tannehill | 5 | 6 | 13 | 6 | 11 | 8 | 8 | 7 | 9 | 9 | 8 | 5 | 8 |
| Matthew Stafford | 8 | 9 | 11 | 13 | 8 | 9 | 9 | 8 | 8 | 8 | 9 | 9 | 9 |
| Mark Sanchez | 22 | 10 | 9 | 9 | 16 | 10 | 10 | 9 | 10 | 10 | 12 | 12 | 12 |
| Russell Wilson | 12 | 13 | 17 | 10 | 10 | 12 | 11 | 12 | 11 | 12 | 11 | 14 | 15 |
| Philip Rivers | 7 | 14 | 15 | 20 | 6 | 17 | 17 | 11 | 16 | 15 | 14 | 10 | 10 |
| Cam Newton | 18 | 12 | 8 | 17 | 19 | 11 | 14 | 14 | 14 | 16 | 21 | 13 | 14 |
| Eli Manning | 17 | – | 18 | 19 | 14 | 19 | 12 | 13 | 12 | 13 | 16 | 23 | 11 |
| Matt Ryan | 21 | 17 | 19 | 15 | 20 | 15 | 15 | 15 | 13 | 11 | 20 | 21 | 13 |
| Andy Dalton | 15 | – | 14 | – | 17 | 14 | 16 | 20 | 15 | 14 | 19 | 22 | 16 |
| Alex Smith | 16 | 11 | 21 | 16 | 18 | 18 | 18 | 16 | 20 | 21 | 13 | 11 | 17 |
| Colin Kaepernick | 11 | 16 | 16 | 11 | 12 | 16 | 21 | 17 | 19 | 18 | 22 | 16 | 21 |
| Joe Flacco | 24 | 15 | 12 | 14 | 24 | 13 | 13 | 18 | 18 | 20 | 15 | 15 | 19 |
| Jay Culter | 13 | 18 | 10 | 12 | 13 | 21 | 19 | 19 | 17 | 17 | 23 | 20 | 18 |
| Josh McCown | 14 | 19 | 22 | 18 | 15 | 22 | 22 | 21 | 21 | 19 | 18 | 17 | 23 |
| Drew Stanton | 20 | 20 | – | 22 | 22 | 20 | 20 | 23 | 22 | 22 | 10 | 19 | 20 |
| Teddy Bridgewater | 23 | 21 | 20 | 21 | 23 | 23 | 23 | 22 | 23 | 24 | 17 | 18 | 22 |
| Brian Hoyer | 19 | – | – | – | 21 | 24 | 24 | 24 | 24 | 23 | 24 | 24 | 24 |

Table 4. NFL starting quarterbacks ranking using BARCW, Borda count, CPDP W-B and CTPDP W-B methods

| Player | Borda | BARCW | CPDP | CTPDP |
|---|---|---|---|---|
| Andrew Luck | 1 | 1 | 1 | 1 |
| Aaron Rodgers | 2 | 2 | 2 | 2 |
| Peyton Manning | 3 | 3 | 3 | 3 |
| Tom Brady | 4 | 4 | 4 | 4 |
| Tony Romo | 5 | 5 | 5 | 5 |
| Drew Brees | 6 | 6 | 6 | 6 |
| Ben Roethlisberger | 7 | 7 | 7 | 7 |
| Ryan Tannehill | 8 | 8 | 8 | 8 |
| Matthew Stafford | 9 | 9 | 9 | 9 |
| Mark Sanchez | 10 | 10 | 10 | 10 |
| Russell Wilson | 11 | 11 | 11 | 11 |
| Philip Rivers | 12 | 12 | 12 | 12 |
| Cam Newton | 13 | 13 | 13 | 14 |
| Eli Manning | 14 | 14 | 14 | 13 |
| Matt Ryan | 15 | 15 | 16 | 15 |
| Joe Flacco | 19 | 16 | 15 | 16 |
| Alex Smith | 17 | 17 | 18 | 19 |
| Colin Kaepernick | 16 | 18 | 17 | 16 |
| Andy Dalton | 20 | 19 | 20 | 16 |
| Jay Cutler | 18 | 20 | 19 | 20 |
| Josh McCown | 21 | 21 | 21 | 21 |
| Drew Stanton | 22 | 22 | 22 | 22 |
| Teddy Bridgewater | 23 | 23 | 23 | 23 |
| Brian Hoyer | 24 | 24 | 24 | 24 |

Table 5. SSE values from Borda count, BARCW, CPDP W-B and CTPDP W-B methods

| Method | Borda | BARCW | CPDP-WB | CTPDP-WB |
|---|---|---|---|---|
| SSE | 2634 | 2590 | 2588 | 2622 |

Figure 1

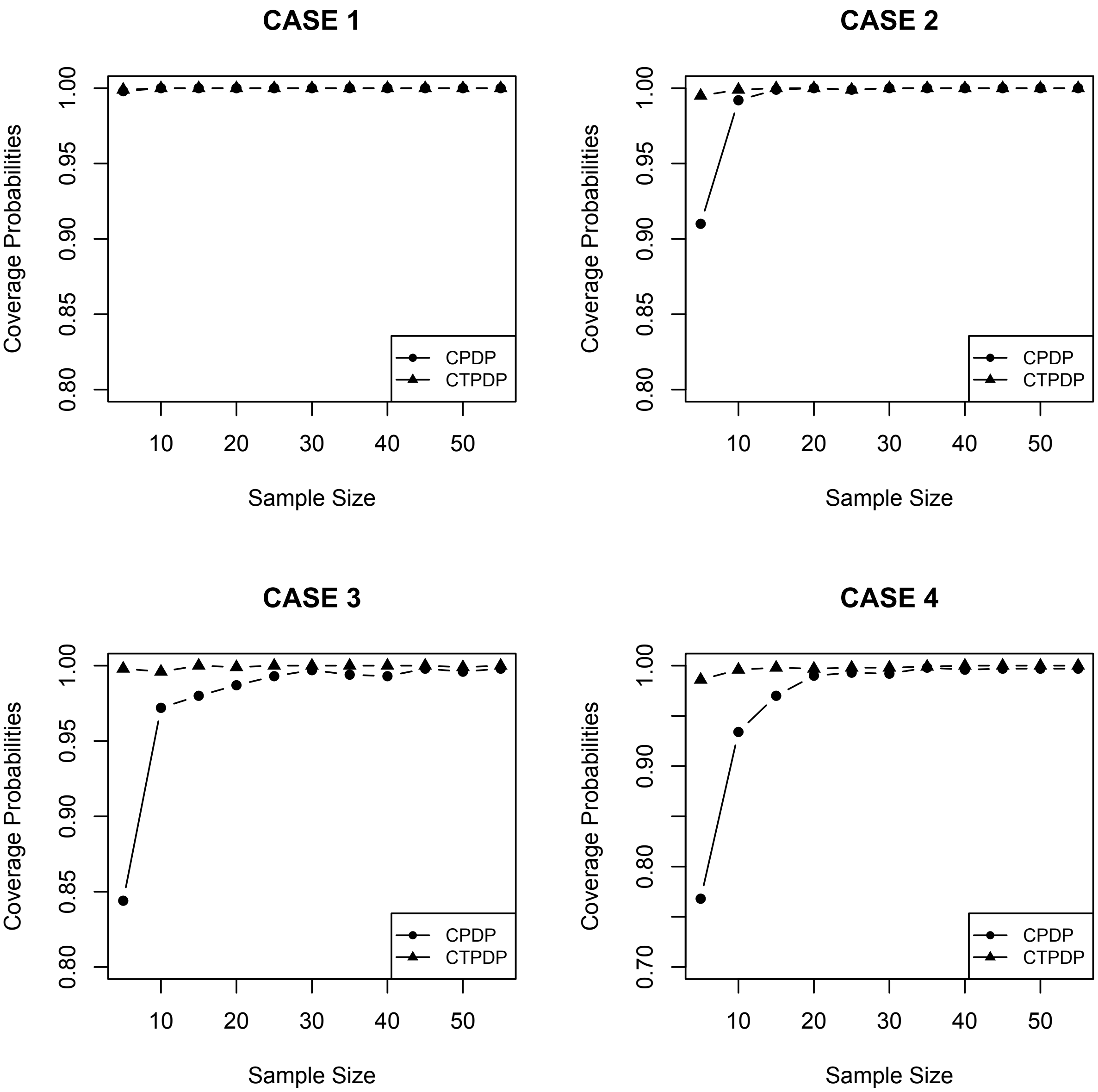

Figure 2

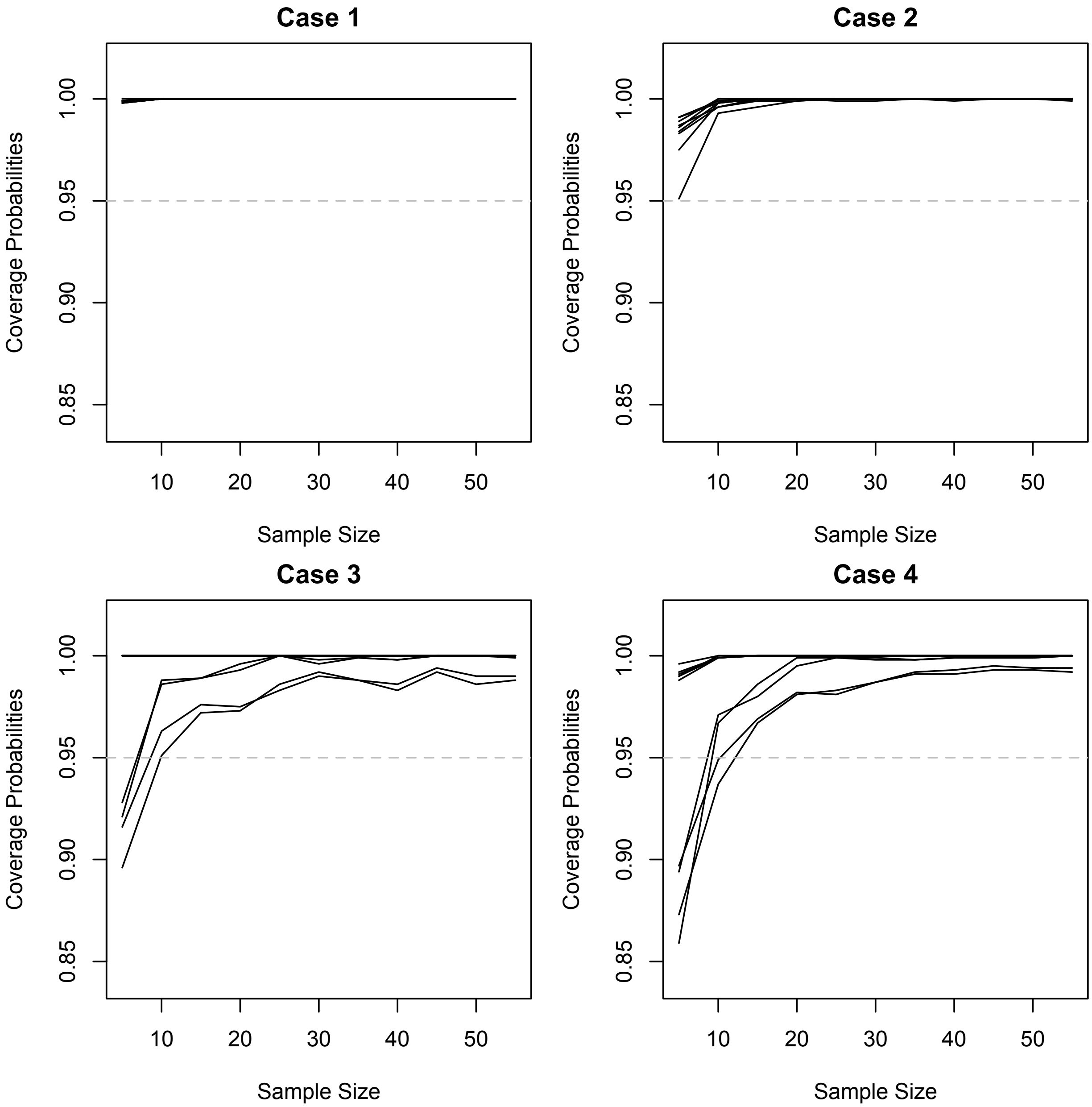

Figure 3

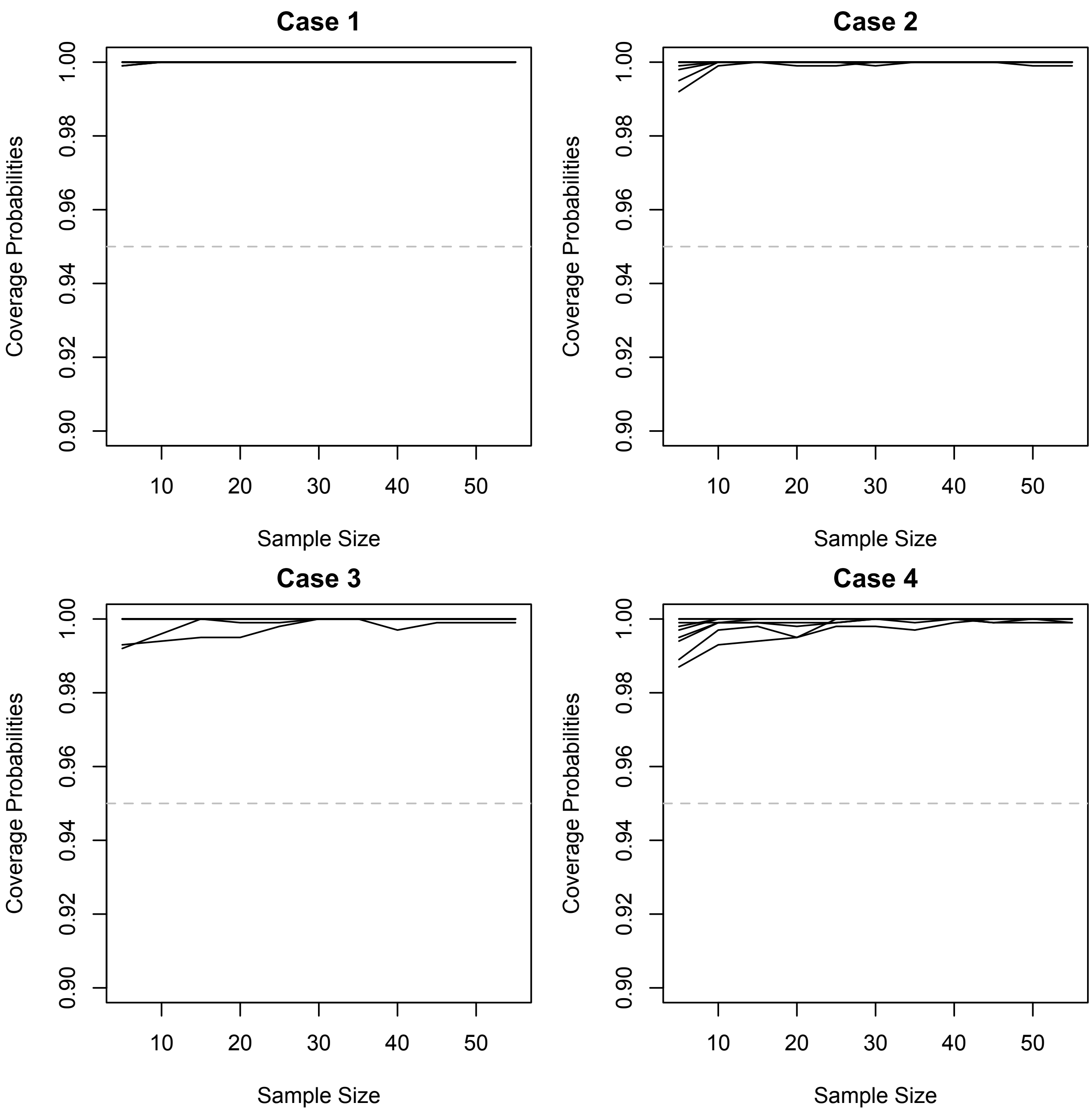

# Figure 4

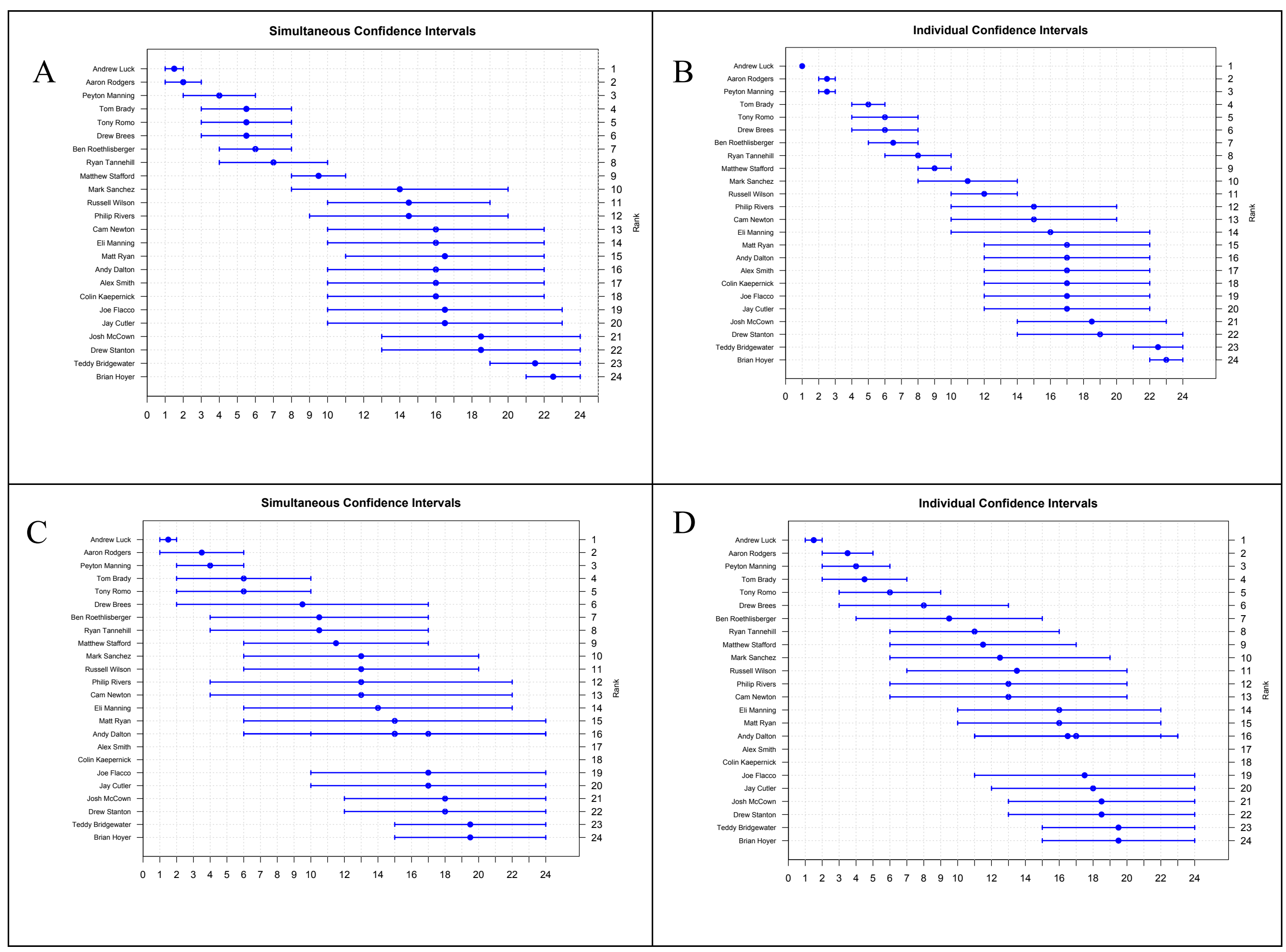

```
# Codes
#=====================================================================
# NFL Pairwise Ranking Analysis: CPDP vs. CTPDP
# Single Run Version (NA-valued Missing Data)
#=====================================================================

rm(list = ls())

nfl <- read.table(
  file = "~/Desktop/nfl1.txt",
  header = FALSE,
  row.names = NULL,
  na.strings = "NA"
)

D <- as.matrix(nfl[, 3:15])

D <- D[rowSums(!is.na(D)) > 0, , drop = FALSE]

N <- nrow(D)
alpha <- 0.05

OBS <- !is.na(D)

cat("Number of teams =", N, "\n")
cat("Number of missing values =", sum(is.na(D)), "\n")

#CPDP method

P_hat <- matrix(0, N, N)
m_ki  <- matrix(0, N, N)

for (k in seq_len(N)) {
  dk <- D[k, ]
  ok <- OBS[k, ]

  for (i in seq_len(N)) {
    overlap <- ok & OBS[i, ]
    m <- sum(overlap)

    if (m > 0) {
      m_ki[k, i] <- m
      diffki <- dk[overlap] - D[i, overlap]
      P_hat[k, i] <- mean(diffki <= 0)
```

```
    }
  }
}

diag(P_hat) <- 1

si_hat <- colSums(P_hat)
ti_hat <- colSums(P_hat > 0.5)

rankCPDP <- rank(si_hat, ties.method = "min")
rankCTPDP <- rank(ti_hat, ties.method = "min")

cat("CPDP ranks:\n")
print(rankCPDP)

cat("CTPDP ranks:\n")
print(rankCTPDP)

Vsp <- numeric(N)

for (i in seq_len(N)) {

  valid_idx <- m_ki[, i] > 0

  V <- sum(
    P_hat[valid_idx, i] *
      (1 - P_hat[valid_idx, i]) /
      m_ki[valid_idx, i]
  )

  S <- 0

  for (s in seq_len(N)) {
    for (t in seq_len(N)) {

      if (s != t) {

        overlap <- OBS[s, ] & OBS[t, ] & OBS[i, ]
        num_sti <- sum(overlap)

        if (num_sti > 0 &&
            m_ki[s, i] > 0 &&
            m_ki[t, i] > 0) {
```

```
          P_sti <- mean(
            ((D[s, overlap] - D[i, overlap]) <= 0) &
            ((D[t, overlap] - D[i, overlap]) <= 0)
          )

          S <- S +
            num_sti *
            (P_sti - P_hat[s, i] * P_hat[t, i]) /
            (m_ki[s, i] * m_ki[t, i])
        }
      }
    }
  }

  Vsp[i] <- max(1e-10, V + S)
}

z_bonf <- qnorm(1 - alpha / (2 * N))

ind_CI1 <- cbind(
  si_hat - z_bonf * sqrt(Vsp),
  si_hat + z_bonf * sqrt(Vsp)
)

sim_CI1 <- matrix(0, N, 2)

for (i in seq_len(N)) {
  sim_CI1[i, 1] <- sum((ind_CI1[i, 1] - ind_CI1[-i, 2]) >= 0) + 1
  sim_CI1[i, 2] <- sum((ind_CI1[i, 2] - ind_CI1[-i, 1]) >= 0) + 1
}

cat("\n CPDP Asymptotical Simultaneous Rank CI\n")
print(cbind(Lower=sim_CI1[,1], Rank=rankCPDP, Upper=sim_CI1[,2]))

CI1_opt <- matrix(0, N, 2)

for (i0 in seq_len(N)) {

  best_width <- Inf

  for (k in seq_len(N)) {

    alpha_vec <- rep((N-k)*alpha/(N*(N-1)), N)
    alpha_vec[i0] <- k*alpha/N
```

```
    z_vec <- abs(qnorm(alpha_vec))

    temp_low <- si_hat - z_vec * sqrt(Vsp)
    temp_up  <- si_hat + z_vec * sqrt(Vsp)

    l_vec <- sapply(seq_len(N),
                    function(x) sum((temp_low[x]-temp_up[-x]) >= 0)+1)

    u_vec <- sapply(seq_len(N),
                    function(x) sum((temp_up[x]-temp_low[-x]) >= 0)+1)

    width <- u_vec[i0] - l_vec[i0]

    if(width < best_width){
      best_width <- width
      CI1_opt[i0,] <- c(l_vec[i0],u_vec[i0])
    }
  }
}

cat("\n CPDP Optimized Individual Rank CI\n")
print(cbind(Lower=CI1_opt[,1], Rank=rankCPDP, Upper=CI1_opt[,2]))

#CTPDP method
Pnki_q <- function(mki,pki,q=0.5){
  if(mki <= 0) return(0)
  cutoff <- floor(mki*q)+1
  pbinom(cutoff-1,size=mki,prob=pki,lower.tail=FALSE)
}

Vstp <- numeric(N)

for(i in seq_len(N)){

  probs_vec <- sapply(seq_len(N),
                      function(k) Pnki_q(m_ki[k,i],P_hat[k,i],0.5))

  V <- sum(probs_vec*(1-probs_vec))

  S <- 0

  for(s in seq_len(N)){
```

```
    for(t in seq_len(N)){
      if(s != t){
        ps <- probs_vec[s]
        pt <- probs_vec[t]
        S <- S + min(ps,pt) - ps*pt
      }
    }
  }

  Vstp[i] <- max(1e-10,V+S)
}

ind_CI2 <- cbind(
  ti_hat - z_bonf*sqrt(Vstp),
  ti_hat + z_bonf*sqrt(Vstp)
)

sim_CI2 <- matrix(0,N,2)

for(i in seq_len(N)){
  sim_CI2[i,1] <- sum((ind_CI2[i,1]-ind_CI2[-i,2]) >= 0)+1
  sim_CI2[i,2] <- sum((ind_CI2[i,2]-ind_CI2[-i,1]) >= 0)+1
}

cat("\n CTPDP Asymptotical Simultaneous Rank CI\n")
print(cbind(Lower=sim_CI2[,1], Rank=rankCTPDP, Upper=sim_CI2[,2]))

CI2_opt <- matrix(0,N,2)

for(i0 in seq_len(N)){

  best_width <- Inf

  for(k in seq_len(N)){

    alpha_vec <- rep((N-k)*alpha/(N*(N-1)),N)
    alpha_vec[i0] <- k*alpha/N

    z_vec <- abs(qnorm(alpha_vec))

    temp_low <- ti_hat - z_vec*sqrt(Vstp)
    temp_up  <- ti_hat + z_vec*sqrt(Vstp)

    l_vec <- sapply(seq_len(N),
```

```
                    function(x) sum((temp_low[x]-temp_up[-x]) >= 0)+1)

    u_vec <- sapply(seq_len(N),
                    function(x) sum((temp_up[x]-temp_low[-x]) >= 0)+1)

    width <- u_vec[i0]-l_vec[i0]

    if(width < best_width){
      best_width <- width
      CI2_opt[i0,] <- c(l_vec[i0],u_vec[i0])
    }
  }
}

cat("\n CTPDP Optimized Individual Rank CI\n")
print(cbind(Lower=CI2_opt[,1], Rank=rankCTPDP, Upper=CI2_opt[,2]))
```

```
#NFL data saved in nfl1.txt

Andrew Luck 1 1 1 3 3 1 1 1 1 1 1 1 1

Aaron Rodgers 2 3 4 2 1 2 3 3 2 2 3 4 3

Peyton Manning 3 2 5 4 2 3 2 2 3 4 4 2 2

Tom Brady 4 7 3 5 4 5 4 6 4 3 6 8 4

Tony Romo 9 5 6 1 5 4 5 4 5 5 7 6 6

Drew Brees 10 4 2 8 9 7 7 5 7 6 2 3 5

Ben Roethlisberger 6 8 7 7 7 6 6 10 6 7 5 7 7

Ryan Tannehill 5 6 13 6 11 8 8 7 9 9 8 5 8

Matthew Stafford 8 9 11 13 8 9 9 8 8 8 9 9 9

Mark Sanchez 22 10 9 9 16 10 10 9 10 10 12 12 12

Russell Wilson 12 13 17 10 10 12 11 12 11 12 11 14 15

Philip Rivers 7 14 15 20 6 17 17 11 16 15 14 10 10

Cam Newton 18 12 8 17 19 11 14 14 14 16 21 13 14

Eli Manning 17  NA 18 19 14 19 12 13 12 13 16 23 11

Matt Ryan 21 17 19 15 20 15 15 15 13 11 20 21 13

Andy Dalton 15 NA 14 NA 17 14 16 20 15 14 19 22 16

Alex Smith 16 11 21 16 18 18 18 16 20 21 13 11 17

Colin Kaepernick 11 16 16 11 12 16 21 17 19 18 22 16 21

Joe Flacco 24 15 12 14 24 13 13 18 18 20 15 15 19

Jay Culter 13 18 10 12 13 21 19 19 17 17 23 20 18
```

Josh McCown 14 19 22 18 15 22 22 21 21 19 18 17 23

Drew Stanton 20 20 NA 22 22 20 20 23 22 22 10 19 20

Teddy Bridgewater 23 21 20 21 23 23 23 22 23 24 17 18 22

Brian Hoyer 19 NA NA NA 21 24 24 24 24 23 24 24 24